\documentclass[dvips]{article}
\usepackage{icrctc07}

\title{H.E.S.S. observations of the supernova remnant RCW 86}
\shorttitle{H.E.S.S. observations of the SNR RCW~86}

\authors{S. Hoppe$^{1}$ \& M. Lemoine-Goumard$^{2}$ for the
H.E.S.S. Collaboration}
\shortauthors{S. Hoppe (for the H.E.S.S. Collaboration) et al.}
\afiliations{$^1$Max-Planck-Institut f\"ur Kernphysik, P.O. Box
103980, Heidelberg, Germany\\ $^2$CENBG, Universit\'e Bordeaux I,
CNRS-IN2P3, Chemin du Solarium, 33175 Gradignan, France} \email{
hoppe@mpi-hd.mpg.de, lemoine@cenbg.in2p3.fr}

\abstract{The shell-type supernova remnant (SNR) RCW 86 - possibly
associated with the historical supernova SN 185 - was observed during
the past three years with the High Energy Stereoscopic System
(H.E.S.S.), an array of four atmospheric-Cherenkov telescopes located
in Namibia. The multi-wavelength properties of RCW 86, e.g. weak radio
emission and North-East X-ray emission almost entirely consisting of
synchroton radiation, resemble those of two very-high energy (VHE;$>$
100 GeV) $\gamma$-ray emitting SNRs RX J1713.7-3946 and RX
J0852-4622. The H.E.S.S. observations reveal a new extended source of
VHE $\gamma$-ray emission.The morphological and spectral properties of
this new source will be presented.}

\begin{document}
\maketitle
\section{Introduction}

Shell-type supernova remnants are widely believed to be the prime
candidates for accelerating cosmic rays up to $10^{15}$~eV. The most
promising way of proving the existence of high energy particles
accelerated in SNR shells is the detection of VHE $\gamma$-rays
produced in nucleonic interactions with ambient matter. Recently, the
H.E.S.S. instrument has detected VHE $\gamma$-ray emission from two
shell-type SNRs, RX~J1713.7-3946 \cite{rxj1713} and RX~J0852.0-4622
\cite{velajr}. They both show an extended morphology highly correlated
with the structures seen in X-rays. Although a hadronic origin is
highly probable in the above cases, a leptonic origin can not be ruled
out.\\ Another young shell-type SNR is RCW~86 (also known as
G315.4-2.3 and MSH14-63). It has a complete shell in
radio~\cite{kesteven}, optical~\cite{smith} and
X-rays~\cite{pisarski}, with a nearly circular shape of $40$'
diameter. It received substantial attention because of its possible
association with SN~185, the first historical galactic
supernova. However, strong evidence for this connection is still
missing: using optical observations, Rosado et al.~\cite{rosado} found
an apparent kinematic distance of 2.8~kpc and an age of ~10 000 years,
whereas recent observations of the North-East part of the remnant with
the Chandra and XMM-Newton satellites strengthen the case that the
event recorded by the Chinese was a supernova and that RCW 86 is its
remnant~\cite{vink}. These observations also reveal that RCW 86 has
properties resembling the already established TeV emitting SNRs
mentioned above: weak radio emission and X-ray emission (almost)
entirely consisting of synchrotron radiation, which could be due to
the expansion of the shock in a wind blown bubble. The South-Western
rim seems to be completely different, with hard X-ray emission,
observed by ROSAT~\cite{bocchino}, mainly coming from stellar ejecta
possibly interacting with circumstellar layers ejected before the SN
explosion. The relatively large size of the remnant - about 40' in
diameter - and the observation of non-thermal X-rays make it a
promising target for $\gamma$-ray observations, aiming at increasing the
currently modest number of remnants where the shells are resolved in
VHE $\gamma$-rays. Hints for $\gamma$-ray emission from RCW 86 were
seen with the CANGAROO-II instrument~\cite{cangaroo}, but no firm
detection was claimed. Here, we present recent data on RCW~86 obtained
with the full H.E.S.S. array in 2005 and 2006 .

\section{The H.E.S.S. detector and the analysis technique}

H.E.S.S. is an array of four imaging Cherenkov telescopes located
1800~m above sea level in the Khomas Highlands in
Namibia~\cite{HESS}. Each telescope has a tesselated mirror with an
area of 107~m$^{2}$~\cite{HESSOptics} and is equipped with a camera
comprising 960 photomultipliers~\cite{HESSCamera} covering a field of
view of 5$^{\circ}$ diameter. Due to the effective rejection of
hadronic showers provided by its stereoscopy, the complete system
(operational since December 2003) can detect point sources at flux
levels of about 1\% of the Crab nebula flux near zenith with a
statistical significance of 5~$\sigma$ in 25 hours of observation
\cite{hess_crab}.  This high sensitivity, the angular resolution of a
few arc minutes and the large field of view make H.E.S.S. ideally
suited for morphology studies of extended VHE $\gamma$-ray sources.\\
The data on RCW~86 were recorded in runs of typically 28 minutes duration in the
so-called ``wobble mode'', where the source is slightly offset from
the center of the field of view. As a cross-check, the obtained data were analysed
using two independent analysis chains, which share only the raw data. 
The first one is based on the combination of a semi-analytical shower model and a
parametrisation based on the moment method of Hillas to yield the
combined likelihood of the event being initiated by a
$\gamma$-ray~\cite{denaurois}. We will call this method the ``Combined
Model analysis'' in the following. The second analysis method is the
standard stereoscopic analysis based on the Hillas parameters of
the shower images~\cite{hillas}.

\section{H.E.S.S. results}
\begin{figure}[h!]
  \begin{minipage}[t]{9.3cm}
    \includegraphics[width=0.8\textwidth]{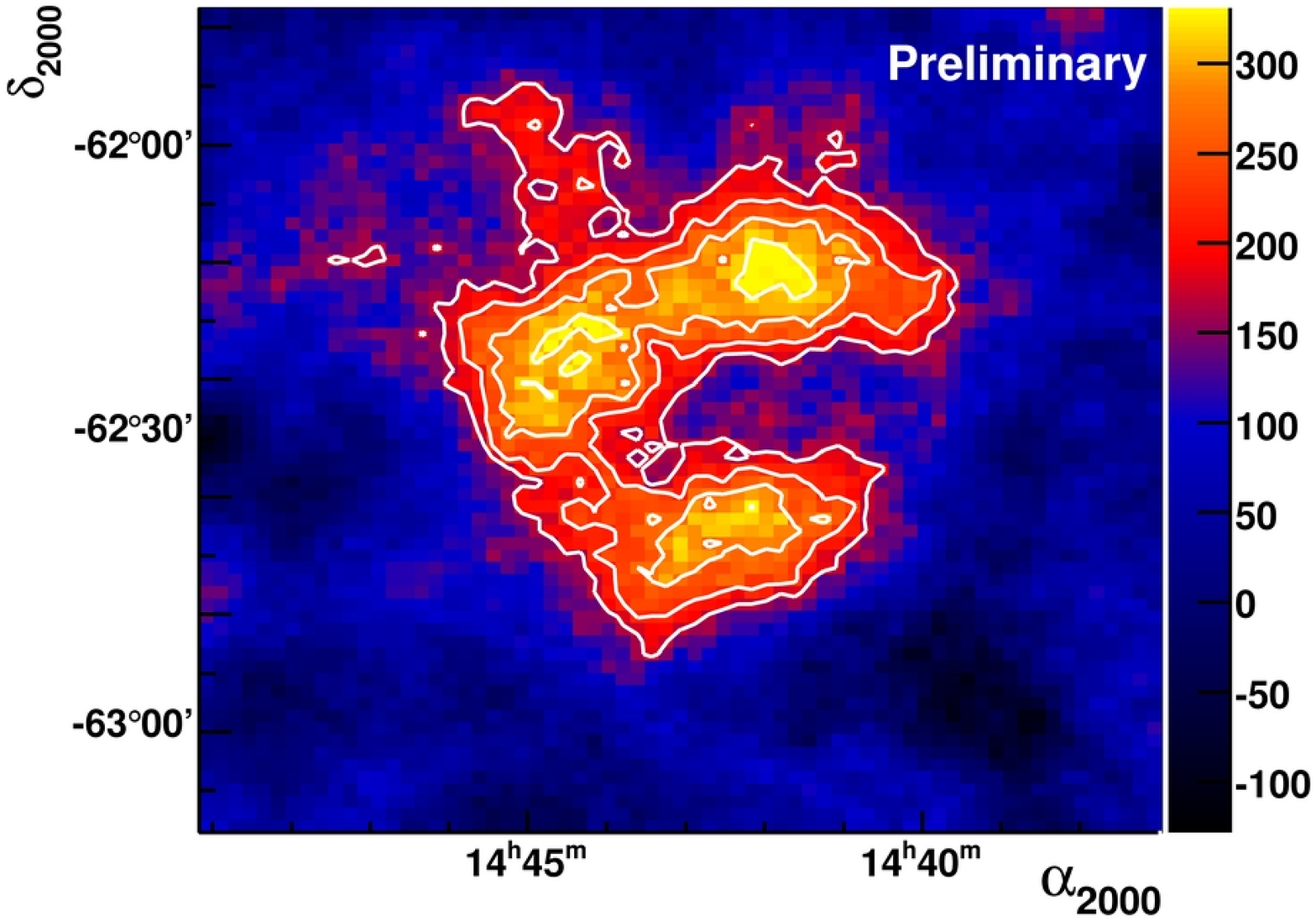}
  \end{minipage}
  \vspace{0.1cm}
  \begin{minipage}[t]{9.3cm}
    \includegraphics[width=0.8\textwidth]{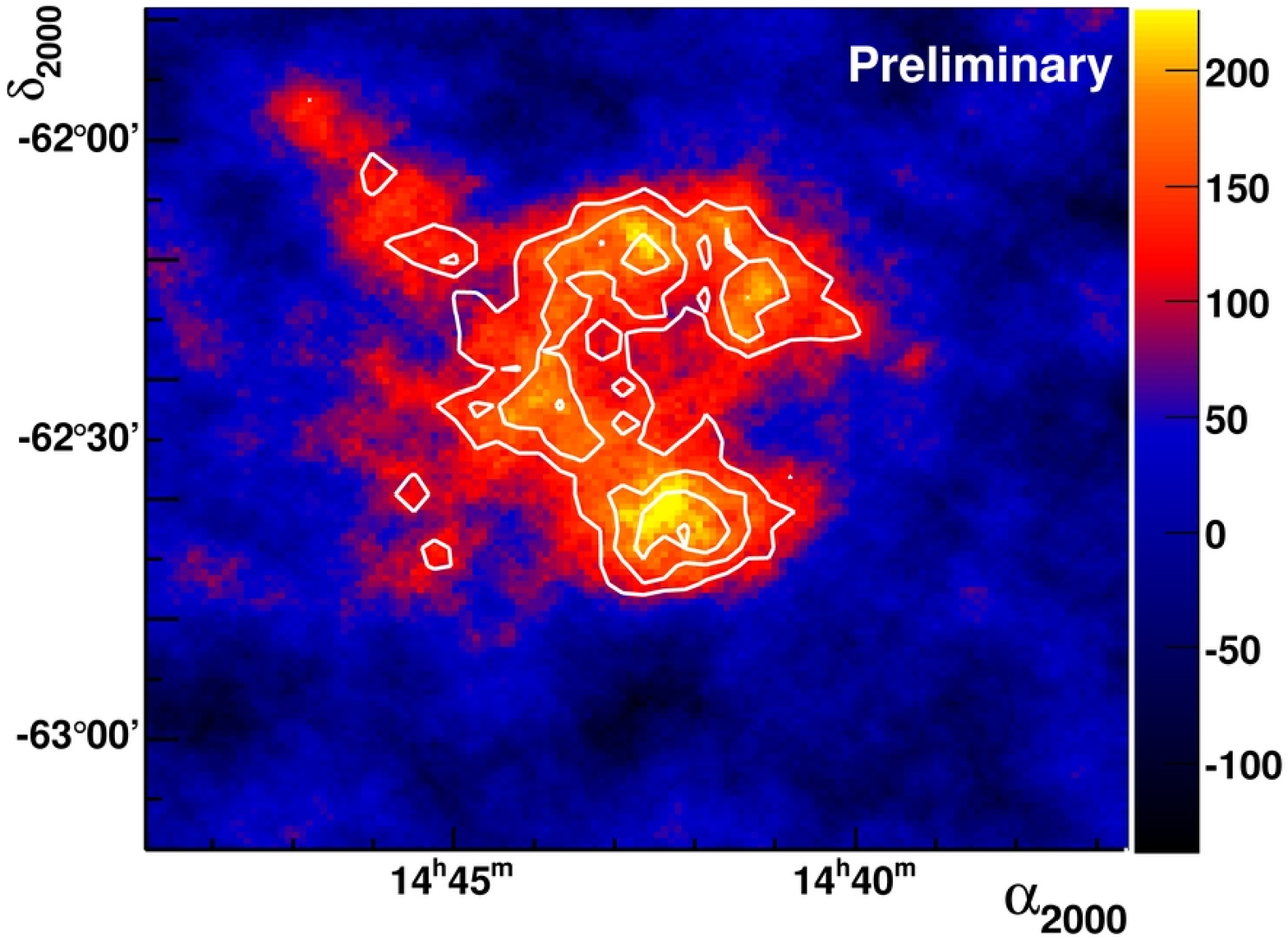}
  \end{minipage}
  \caption{VHE $\gamma$-ray emission from RCW~86, as seen
with H.E.S.S.. The top image shows the excess skymap obtained with the
Combined Model analysis where shower images are matched against image
templates, whereas the bottom image results from the classical,
slightly less sensitive Hillas analysis technique. White contours
correspond to 3, 4, 5, 6 sigma significance, obtained by counting
gamma rays within 0.11$^{\circ}$ from a given location.
\label{fig:rcw86model}}
\end{figure}

\begin{figure}[t!]
\centering
\includegraphics[width=0.4\textwidth]{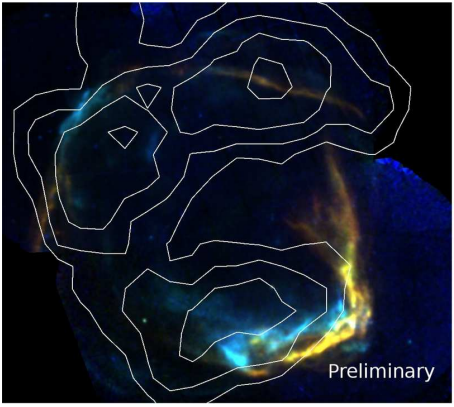}
\caption{Significance contours of gamma-ray emission (from the Combined Model analysis; 3, 4, 5, 6 sigma)
superimposed onto the XMM X-ray image of the remnant \cite{vink}.
\label{fig:xmm}}
\end{figure}

RCW 86 was observed for about 30 hours with the H.E.S.S. instrument
with a mean zenith angle of $41^{\circ}$. Within a circular region
of 27' radius (defined a priori so that it encompasses the
whole remnant) around the centre of the SNR ($\alpha_{J2000}$ =
14$^h$42$^m$43$^s$, $\delta_{J2000}$ = $-62^\circ$29'), a clear VHE
$\gamma$-ray signal with more than $9$ standard deviations is detected
with both analysis methods described above. The exact morphology of the
gamma-ray emission is still under study: whereas one type of data
analysis shows hints of a 3/4 shell resembling the shape of the X-ray
emission (Fig.~\ref{fig:rcw86model} top, Fig~\ref{fig:xmm}), this
morphology is not quite as evident with the other analysis technique
(Fig.~\ref{fig:rcw86model} bottom), and more data may be required to
fully settle this issue. The differential energy spectrum of RCW~86, $\phi(E)$,
was extracted from a circular region of diameter 22' around the position 
$\alpha_{J2000}$ = 14$^h$42$^m$12$^s$, $\delta_{J2000}$ = $-62^\circ$24' 
which is -- different from the region for which the detection significance was
determined -- adjusted to the H.E.S.S. data to include $\sim$ 90 \% of
the $\gamma$-ray excess. It is well described by a power-law with a spectral index of $\Gamma = 2.5 \pm 0.1_{\rm{stat}}$ and a flux normalisation at 1 TeV
of $\phi(1TeV) = (2.71 \pm 0.35_{\rm{stat}}) \times 10^{-12} \mathrm{cm^{-2}}
\mathrm{s^{-1}} \mathrm{TeV^{-1}} $ . The integral flux in the energy
range 1 - 10 TeV is $\sim$ 8\% of the integrated flux
of the Crab nebula within the same range. However, at this level of data
statistics, a power-law with index $\Gamma \sim 1.9$ and an
exponential cut-off at $E_c^{\gamma} \sim 5$ TeV is also a good fit to
the data.

\section{Discussion}
There are two basic mechanisms for $\gamma$-ray production in young
supernova remnants, inverse Compton scattering of high energy
electrons off ambient photons (leptonic scenario) and $\pi^0$ mesons
produced in inelastic interactions of accelerated protons with ambient
gas decaying into $\gamma$-rays (hadronic scenario). The measured
$\gamma$-ray flux spectrum from RCW~86 translates into an energy flux
between 2 and 10 TeV of $3.4 \times 10^{-12} \, \rm erg \, cm^{-2} \,
s^{-1}$. In a leptonic scenario, the ratio of this energy flux and the
X-ray flux between 2 and 10 keV ($1.7 \times 10^{-10} \, \rm erg \,
cm^{-2} \, s^{-1}$, see Winkler \cite{winkler}) determines the magnetic field to be
close to $22 \, \mu \rm G$. This value, completely independent of the
distance and age of the SNR, is compatible with the calculation made
by J. Vink et al.~\cite{vink} based on the thin filaments resolved by
Chandra for a distance of 2.5~kpc in which he also deduces a high
speed of the blast wave ($\sim 2700 \, \rm km s^{-1}$). However, it is
more than ten times lower than the value proposed by H. J. Voelk and
his colleagues~\cite{volk} for the same distance using a much lower
velocity of the shock of $800 \, \rm km s^{-1}$ as suggested by
optical data in the Southern region of the SNR~\cite{rosado}.\\ In a
hadronic scenario, one can estimate the total energy in accelerated
protons $W_p$ in the range $10 - 100$~TeV required to produce the
$\gamma$-ray luminosity $L_{\gamma}$ observed by H.E.S.S. using the
relation:
\begin{equation}
W_p (10 - 100 \mathrm{TeV}) \approx \tau_{\gamma} \times L_{\gamma}(1
- 10 \mbox{TeV})
\end{equation}
in which $\tau_{\gamma} \approx 4.4 \times 10^{15} \left(\frac{n}{1 \,
\mathrm{cm^{-3}}} \right)^{-1}$ is the characteristic cooling time of
protons through the $\pi^0$ production channel. The correspnding
$L_{\gamma}$ can be calculated using:
\begin{eqnarray}
  L_{\gamma} (1-10 \, \mathrm{TeV}) & = & 4 \pi D^2
  \int_{1\,\mathrm{TeV}}^{10\,\mathrm{TeV}} E \phi(E) dE \nonumber \\
  & = & 2.8 \times 10^{31} \big(\frac{D}{200 \, \mathrm{pc}} \big)^2
  \mathrm{erg \, s^{-1}} \nonumber
\end{eqnarray} 
Finally, the total energy injected in protons is calculated by
extrapolating the proton spectrum with the same spectral shape as the
photon spectrum down to 1 GeV. Therefore, this estimation is highly
dependent on the shape of the $\gamma$-ray spectrum. Assuming that the
proton spectrum is a power-law with index $\Gamma = 2.5$, one would
obtain a total energy injected into protons of $W_p \rm (tot) = 3
\times 10^{51} \left( \frac{D}{2.5 \, \mathrm{kpc}} \right)^2 \left(
\frac{n}{1 \, \mathrm{cm^{-3}}} \right)^{-1} \, \rm erg$. 
For densities of $\sim 1 \rm \, cm^{-3}$, the only way to explain the
entire $\gamma$-ray flux by proton-proton interactions in a
homogeneous medium is to assume that RCW~86 is a close supernova
remnant ($\sim 1$ kpc). Indeed, for larger distances and a typical
energy of the supernova explosion of $10^{51}$~erg, the acceleration
efficiency would be excessive. For an exponential cut-off power-law
with $\Gamma= 1.9$ and $E_c = 10 \times E_c^{\gamma} = 50$ TeV, the
total energy injected into protons would be $10^{50} \left(
\frac{D}{2.5 \, \mathrm{kpc}} \right)^2 \left( \frac{n}{1 \,
\mathrm{cm^{-3}}} \right)^{-1}$ erg which would make the hadronic
scenario possible even at larger distances. However, the observation
of TeV gamma-rays from the remnant up to more than 10 TeV favors
somewhat the scenario of a young -- and therefore close-by -- remnant
with high expansion speed, easing the acceleration of high-energy
particles.

\section{Summary}

H.E.S.S. observations have led to the discovery of the shell-type SNR
RCW~86 in VHE $\gamma$-rays . The $\gamma$-ray signal is extended but
the exact morphology of the emission region is still under
study. The flux from the remnant is $\sim$8\% of the flux from the
Crab nebula, with a similar spectral index of 2.5, but the spectrum is
also well described by a power law with index 1.9 and a cutoff around
5 TeV. The question of the nature of the particles producing the
$\gamma$-ray signal observed by H.E.S.S. was also addressed. However,
at present, no firm conclusions can be drawn from the spectral shape.

\section{Acknowledgements}
The support of the Namibian authorities and of the University of Namibia
in facilitating the construction and operation of H.E.S.S. is gratefully
acknowledged, as is the support by the German Ministry for Education and
Research (BMBF), the Max Planck Society, the French Ministry for Research,
the CNRS-IN2P3 and the Astroparticle Interdisciplinary Programme of the
CNRS, the U.K. Science and Technology Facilities Council (STFC),
the IPNP of the Charles University, the Polish Ministry of Science and 
Higher Education, the South African Department of
Science and Technology and National Research Foundation, and by the
University of Namibia. We appreciate the excellent work of the technical
support staff in Berlin, Durham, Hamburg, Heidelberg, Palaiseau, Paris,
Saclay, and in Namibia in the construction and operation of the
equipment.

\bibliography{icrc0280}

\begin{thebibliography}{10}

\bibitem{velajr}
F.~{Aharonian et al.}
\newblock {\em A\&A}, 437:L7--L10, 2005.

\bibitem{hess_crab}
F.~{Aharonian et al.}
\newblock {\em A\&A}, 457:899--915, 2006.

\bibitem{rxj1713}
F.~{Aharonian et al.}
\newblock {\em A\&A}, 464:235--243, 2007.

\bibitem{hillas}
F.~{Aharonian et al. ({\it H.E.S.S. Collaboration})}.
\newblock {\em A\&A}, 430:865, 2005.

\bibitem{HESSOptics}
K.~{Bernl\"ohr et al.}
\newblock {\em APh}, 20:111, 2003.

\bibitem{bocchino}
F.~{Bocchino et al.}
\newblock {\em A\&A}, 360:671, 2000.

\bibitem{denaurois}
M.~{de Naurois}.
\newblock In {\em Proceedings of ``Towards a Network of Atmospheric Cherenkov
  Detectors VII''}, 2005.

\bibitem{HESS}
J.~A. {Hinton}.
\newblock {\em NewAR}, 48:331, 2004.

\bibitem{kesteven}
M.~J. {Kesteven} and J.~L. {Caswell}.
\newblock {\em A\&A}, 183:118, 1987.

\bibitem{pisarski}
P.~L. {Pisarski et al.}
\newblock {\em ApJ}, 277:710, 1984.

\bibitem{rosado}
M.~{Rosado et al.}
\newblock {\em A\&A}, 315:243, 1996.

\bibitem{smith}
R.~C. {Smith}.
\newblock {\em AJ}, 114:2664, 1997.

\bibitem{HESSCamera}
P.~{Vincent ({\it H.E.S.S. Collaboration})}.
\newblock In {\em Proceedings of the 28th ICRC, T. Kajita et al., Eds.
  (Universal Academy Press)}, page 2887, 2003.

\bibitem{vink}
J.~{Vink et al.}
\newblock {\em ApJL}, 648:33, 2006.

\bibitem{volk}
H.~J. {V\"olk et al.}
\newblock {\em A\&A}, 433:229, 2005.

\bibitem{cangaroo}
S.~{Watanabe ({\it CANGAROO Collaboration})}.
\newblock In {\em Proceedings of the 28th ICRC, T. Kajita et al., Eds.
  (Universal Academy Press)}, 2003.

\bibitem{winkler}
P.~F. {Winkler Jr.}
\newblock {\em ApJ}, 221:220, 1978.

\end{thebibliography}
\bibliographystyle{plain}
\end{document}